\documentclass[aps,prl,preprint,tightenlines,superscriptaddress,showpacs]{revtex4}

\usepackage{graphicx}

\newcommand{\ACP}{A_{\mathit{CP}}}
\newcommand{\ACPraw}{A_{\mathit{CP}}^{\mathrm{raw}}}
\newcommand{\BB}{B\bar{B}}
\newcommand{\BF}{\mathcal{B}}
\newcommand{\bdgamma}{b \to d\gamma}
\newcommand{\bsgamma}{b \to s\gamma}
\newcommand{\CP}{\mathit{CP}}

\newcommand{\DE}{\Delta E}
\newcommand{\Ebeam}{E_{\mathrm{beam}}^{*}}
\newcommand{\Egamma}{E_\gamma^*}

\newcommand{\fbi}{\mathrm{fb}^{-1}}
\newcommand{\GeV}{\mathrm{GeV}}

\newcommand{\Mbc}{M_{\mathrm{bc}}}
\newcommand{\MeV}{\mathrm{MeV}}
\newcommand{\MXs}{M_{X_s}}
\newcommand{\KS}{K^0_S}
\newcommand{\qq}{q\bar{q}}

\newcommand{\LumONRES}{140}
\newcommand{\LumOFFRES}{15}
\newcommand{\NBBmillunit}{152}

\newcommand{\GaussLambdaAdderrratio}{23\%}

\newcommand{\BFbsgammaTheory}{(3.57 \pm 0.30) \times 10^{-4}}

\newcommand{\ACPminCLEO}{-0.27}
\newcommand{\ACPmaxCLEO}{0.10}

\newcommand{\effKID}{90\%}
\newcommand{\fakeKID}{7\%}
\newcommand{\effPIID}{93\%}

\newcommand{\MCeffLeptonSig}{15\%}
\newcommand{\MCrejLeptonQQ}{98.5\%}
\newcommand{\MCeffLRSig}{92\%}
\newcommand{\MCrejLRQQ}{55\%}

\newcommand{\wrongtagOne}{0.0206 \pm 0.0027} 
\newcommand{\wrongtagTwo}{0.248 \pm 0.020} 
\newcommand{\wrongtagThree}{0.0067 \pm 0.0013} 
\newcommand{\dilution}{1.041 \pm 0.006} 
\newcommand{\YieldMinusNosyst}{393.2 \pm 25.9}
\newcommand{\YieldPlusNosyst}{392.0 \pm 25.9}
\newcommand{\YieldZeroNosyst}{52.8 \pm 9.6}
\newcommand{\YBBFixedTagged}{39.3}
\newcommand{\YBBFixedAmbiguous}{9.2}
\newcommand{\YrareBFixedTagged}{35.4}
\newcommand{\YrareBFixedAmbiguous}{3.3}
\newcommand{\wrongtagOneDpi}{0.0314 \pm 0.0039}
\newcommand{\wrongtagTwoDpi}{0.285 \pm 0.024}
\newcommand{\wrongtagThreeDpi}{0.0225 \pm 0.0033}
\newcommand{\dilutionDpi}{1.059 \pm 0.013}
\newcommand{\YieldPlusDpi}{2105 \pm 60}
\newcommand{\YieldZeroDpi}{580 \pm 30}
\newcommand{\YieldMinusDpi}{2125 \pm 60}
\newcommand{\ACPDpi}{0.006 \pm 0.022}
\newcommand{\ACPnosyst}{0.002 \pm 0.050}

\newcommand{\ACPerrorFit}{0.014}

\newcommand{\ACPerrorRareB}{0.014}
\newcommand{\ACPerrorDetector}{0.022}
\newcommand{\ACPmaxerrorbdgamma}{0.001}
\newcommand{\ACPsyst}{0.030}

\newcommand{\ACPfinalSTR}{%
  \ACPnosyst \mbox{(stat.)} \pm \ACPsyst \mbox{(syst.)}}
\newcommand{\ACPmin}{-0.093}
\newcommand{\ACPmax}{0.096}

\begin{document}

\vspace*{-3\baselineskip}
\resizebox{!}{3cm}{\includegraphics{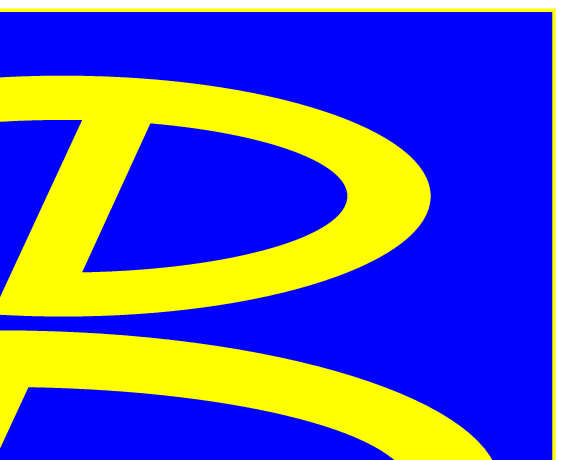}}

\preprint{Belle Preprint 2004-12}
\preprint{KEK Preprint 2004-6}

\title{Measurement of the $\CP$ Asymmetry in $B \to X_s\gamma$}


\affiliation{Budker Institute of Nuclear Physics, Novosibirsk}
\affiliation{Chiba University, Chiba}
\affiliation{Chonnam National University, Kwangju}
\affiliation{University of Cincinnati, Cincinnati, Ohio 45221}
\affiliation{University of Frankfurt, Frankfurt}
\affiliation{Gyeongsang National University, Chinju}
\affiliation{University of Hawaii, Honolulu, Hawaii 96822}
\affiliation{High Energy Accelerator Research Organization (KEK), Tsukuba}
\affiliation{Hiroshima Institute of Technology, Hiroshima}
\affiliation{Institute of High Energy Physics, Chinese Academy of Sciences, Beijing}
\affiliation{Institute of High Energy Physics, Vienna}
\affiliation{Institute for Theoretical and Experimental Physics, Moscow}
\affiliation{J. Stefan Institute, Ljubljana}
\affiliation{Kanagawa University, Yokohama}
\affiliation{Korea University, Seoul}
\affiliation{Kyungpook National University, Taegu}
\affiliation{Swiss Federal Institute of Technology of Lausanne, EPFL, Lausanne}
\affiliation{University of Ljubljana, Ljubljana}
\affiliation{University of Maribor, Maribor}
\affiliation{University of Melbourne, Victoria}
\affiliation{Nagoya University, Nagoya}
\affiliation{Nara Women's University, Nara}
\affiliation{National United University, Miao Li}
\affiliation{Department of Physics, National Taiwan University, Taipei}
\affiliation{H. Niewodniczanski Institute of Nuclear Physics, Krakow}
\affiliation{Nihon Dental College, Niigata}
\affiliation{Niigata University, Niigata}
\affiliation{Osaka City University, Osaka}
\affiliation{Osaka University, Osaka}
\affiliation{Panjab University, Chandigarh}
\affiliation{Peking University, Beijing}
\affiliation{Princeton University, Princeton, New Jersey 08545}
\affiliation{RIKEN BNL Research Center, Upton, New York 11973}
\affiliation{Saga University, Saga}
\affiliation{University of Science and Technology of China, Hefei}
\affiliation{Seoul National University, Seoul}
\affiliation{Sungkyunkwan University, Suwon}
\affiliation{University of Sydney, Sydney NSW}
\affiliation{Tata Institute of Fundamental Research, Bombay}
\affiliation{Toho University, Funabashi}
\affiliation{Tohoku Gakuin University, Tagajo}
\affiliation{Tohoku University, Sendai}
\affiliation{Department of Physics, University of Tokyo, Tokyo}
\affiliation{Tokyo Institute of Technology, Tokyo}
\affiliation{Tokyo Metropolitan University, Tokyo}
\affiliation{Tokyo University of Agriculture and Technology, Tokyo}
\affiliation{University of Tsukuba, Tsukuba}
\affiliation{Virginia Polytechnic Institute and State University, Blacksburg, Virginia 24061}
\affiliation{Yokkaichi University, Yokkaichi}
\affiliation{Yonsei University, Seoul}
  \author{S.~Nishida}\affiliation{High Energy Accelerator Research Organization (KEK), Tsukuba} 
  \author{K.~Abe}\affiliation{High Energy Accelerator Research Organization (KEK), Tsukuba} 
  \author{K.~Abe}\affiliation{Tohoku Gakuin University, Tagajo} 
  \author{T.~Abe}\affiliation{High Energy Accelerator Research Organization (KEK), Tsukuba} 
  \author{H.~Aihara}\affiliation{Department of Physics, University of Tokyo, Tokyo} 
  \author{Y.~Asano}\affiliation{University of Tsukuba, Tsukuba} 
  \author{T.~Aushev}\affiliation{Institute for Theoretical and Experimental Physics, Moscow} 
  \author{S.~Bahinipati}\affiliation{University of Cincinnati, Cincinnati, Ohio 45221} 
  \author{A.~M.~Bakich}\affiliation{University of Sydney, Sydney NSW} 
  \author{Y.~Ban}\affiliation{Peking University, Beijing} 
% ...
\collaboration{The Belle Collaboration}

\noaffiliation


\begin{abstract}
Direct $\CP$ violation in the $\bsgamma$ process is a sensitive probe
of physics beyond the Standard Model.
We report a measurement of the $\CP$ asymmetry
in $B \to X_s\gamma$, where the hadronic recoil system
$X_s$ is reconstructed using a pseudo-reconstruction technique.
In this approach there is negligible contamination from $\bdgamma$ decays,
which are expected to have a much larger $\CP$ asymmetry.
We find $\ACP = \ACPfinalSTR$ for $B \to X_s\gamma$ events
having recoil mass smaller than $2.1~\GeV/c^2$.
The analysis is based on a data sample of $\LumONRES~\fbi$ recorded at the
$\Upsilon(4S)$ resonance with the Belle detector at the KEKB $e^+e^-$
storage ring.
\end{abstract}


\maketitle


Radiative $B$ decays,
which proceed mainly through the $\bsgamma$ process,
have played an important role in the search for physics beyond
the Standard Model (SM).
The inclusive branching fraction has been measured by
CLEO~\cite{Chen:2001fj}, ALEPH~\cite{Barate:1998vz} and
Belle~\cite{Abe:2001hk},
giving results consistent with the recent theoretical prediction
of $\BFbsgammaTheory$~\cite{Buras:2002tp}.
The SM predicts very small direct $\CP$ violation in $\bsgamma$;
the $\CP$ asymmetry,
\[
 \ACP(\bsgamma)
 = \frac{\Gamma(\bsgamma) - \Gamma(\bar{b} \to \bar{s}\gamma)}{%
 \Gamma(\bsgamma) + \Gamma(\bar{b} \to \bar{s}\gamma)}\, ,
\]
is about $+0.5\%$ in the SM,
while some new physics, such as supersymmetry,
allows the $\CP$ asymmetry to be above $10\%$
without changing the inclusive branching fraction~\cite{ACP-predictions}.
Thus, measurement of $\ACP$
may provide information on new physics
that cannot be extracted from the measurement
of the branching fraction.
Previously, CLEO measured
$\ACPminCLEO < 0.965 \ACP(\bsgamma) + 0.02 \ACP(\bdgamma) < \ACPmaxCLEO$
at $90\%$ confidence level~\cite{Coan:2000pu}; however
$\bdgamma$ is expected to cancel
the $\CP$ asymmetry in $\bsgamma$~\cite{Akeroyd:2001gf}.
The measurement presented here has negligible $\bdgamma$ contamination.

In this paper, we report on a measurement of the $\CP$
asymmetry in $B \to X_s\gamma$,
where the hadronic recoil system $X_s$ is reconstructed
using a pseudo-reconstruction technique.
The analysis is based on
$\LumONRES~\fbi$ of data taken at the $\Upsilon(4S)$ resonance
(on-resonance) and
$\LumOFFRES~\fbi$ at an energy $60~\MeV$ below the resonance
(off-resonance),
which was recorded by the Belle detector~\cite{Mori:2000cg}
at the KEKB asymmetric $e^+e^-$ collider
($3.5~\GeV$ on $8~\GeV$)~\cite{KEKB:NIM}.
The on-resonance data correspond to $\NBBmillunit$ million $\BB$ events.
The Belle detector
has a three-layer silicon vertex detector,
a 50-layer central drift chamber (CDC), an array of aerogel
Cherenkov counters (ACC), time-of-flight scintillation counters (TOF)
and an electromagnetic calorimeter of CsI(Tl) crystals (ECL) located inside
a superconducting solenoid coil that provides a 1.5 T magnetic field.
An instrumented iron flux-return for $K_L$/$\mu$ detection
is located outside the coil.

We form the hadronic recoil system, $X_s$, with a mass up to $2.1~\GeV/c^2$,
by combining
one charged or neutral kaon with one to four pions,
where at most one pion can be neutral.
We also reconstruct $X_s$
via $K^{\pm}K^{\mp}K^{\pm}(\pi^{\mp})$
and $\KS K^{\pm}K^{\mp}(\pi^{\pm})$
including the $B \to K\phi\gamma$ decays
that were observed recently by Belle~\cite{Drutskoy:2003xh}.

Each of the primary charged tracks is
required to have
a momentum in the $e^+e^-$ center-of-mass (CM) frame that is greater 
than $100~\MeV/c$, and to have an impact parameter
within $\pm 5 \mathrm{~cm}$
of the interaction point along the positron beam axis
and within $0.5 \mathrm{~cm}$ in the transverse plane.
These tracks are identified as pion or kaon candidates
according to a likelihood ratio based on the light yield in the ACC,
the TOF information and the specific ionization measurements in the CDC.
For the selection applied on the likelihood ratio,
we obtain an efficiency
(pion misidentification probability) of $\effKID$ ($\fakeKID$)
for charged kaon candidates,
and an efficiency
(kaon misidentification probability) of $\effPIID$ ($\fakeKID$)
for charged pion candidates.
Tracks that are identified as an electron or muon are excluded.

$\KS$ candidates are formed from $\pi^+\pi^-$ combinations
whose invariant mass is within $8~\MeV/c^2$
of the nominal $\KS$ mass.
The two pions are required to have a common vertex that
is displaced from the interaction point.
The $\KS$ momentum direction is also required to be
consistent with the $\KS$ flight direction.
Neutral pion candidates are formed from pairs of photons
that have an invariant mass within $16~\MeV/c^2$ of
the nominal $\pi^0$ mass and an opening angle smaller than $60^\circ$.
Each photon is required to have an energy greater than $50~\MeV$.
A mass constrained fit is then performed to obtain the $\pi^0$ momentum.

The $B$ meson candidates are reconstructed from
the $X_s$ system and the highest energy photon
with a CM energy between $1.8$ and $3.4~\GeV$
within the acceptance of the barrel ECL
($33^\circ<\theta_\gamma<128^\circ$, where $\theta_\gamma$
is the polar angle of the photon in the laboratory frame).
In order to reduce the background from
decays of $\pi^0$ and $\eta$ mesons,
we combine the photon candidate
with all other photons in the event
and reject the event if the invariant mass of any pair is
within $18~\MeV/c^2$ ($32~\MeV/c^2$)
of the nominal $\pi^0$ ($\eta$) mass.

We use two independent kinematic variables for the $B$ reconstruction:
the beam-energy constrained mass
$\Mbc \equiv \sqrt{\left(\Ebeam/c^2\right)^2
  - (|\vec{p}_{X_s}^{\,*}+\vec{p}_\gamma^{\,*}|/c)^2}$
and
$\DE \equiv E_{X_s}^* + \Egamma - \Ebeam$,
where $\Ebeam$ is the beam energy,
and $\vec{p}_\gamma^{\,*}$, $\Egamma$,
$\vec{p}_{X_s}^{\,*}$, $E_{X_s}^*$ are
the momenta and energies of the photon
and the $X_s$ system, respectively, calculated in the CM frame.
In the $\Mbc$ calculation, the photon momentum is rescaled
so that $|\vec{p}_\gamma^{\,*}|=(\Ebeam-E_{X_s}^*)/c$
is satisfied; this improves the $\Mbc$ resolution to $2.9~\MeV/c^2$.
We require $\Mbc > 5.24~\GeV/c^2$ and $-150~\MeV < \DE < 80~\MeV$.
We define the signal region to be $\Mbc > 5.27~\GeV/c^2$.

The largest source of background originates from
continuum $e^+e^- \to \qq$ ($q = u,d,s,c$) production
including contributions from
initial state radiation ($e^+e^-\to q\bar{q}\gamma$).
In order to suppress this background,
we require the presence of a high energy lepton from the opposite $B$.
The lepton can be either an electron with a CM momentum greater
than $0.8~\GeV/c$ or a muon with a laboratory momentum greater
than $1.0~\GeV/c$.
In both cases, the opening angle $\theta^*_{\ell\gamma}$
between the high energy photon from the signal $B$ meson
and the lepton, calculated in the CM frame,
must satisfy $|\cos\theta^*_{\ell\gamma}| < 0.8$.
In addition to this lepton requirement,
we use the likelihood ratio described in Ref.~\cite{Nishida:2002me},
which utilizes the information from
a Fisher discriminant~\cite{Fisher:1936et}
formed from six modified Fox-Wolfram moments~\cite{Fox:1978vu}
and the cosine of the angle between
the $B$ meson flight direction and the beam axis.
The lepton requirement (likelihood ratio selection)
retains $\MCeffLeptonSig$ ($\MCeffLRSig$) of the signal,
rejecting $\MCrejLeptonQQ$ ($\MCrejLRQQ$) of the $\qq$ background.

When multiple $B$ candidates are found in the same event,
we take the candidate that gives the highest confidence level
when we fit the $X_s$ decay vertex,
constrained to the profile of the measured interaction region.
The confidence level for the $\KS\pi^0\gamma$ mode is
set to zero, because we do not determine the vertex.
If multiple $B$ candidates appear in an event according to
the inclusion or omission of different $\pi^0$ mesons in the $X_s$
recoil system, we take the candidate that has the minimum $|\DE|$.

We fit the $\Mbc$ distribution to extract the signal yield.
The $\Mbc$ distribution of the $\qq$ background is
modeled by an ARGUS function~\cite{Albrecht:1990am}
whose shape is determined from the off-resonance data.
(The lepton requirement is not applied to the off-resonance data
in order to compensate for the limited amount of data in that sample.)
Background from $B$ decay is divided into two components.
$B$ decays through $b \to c$ transitions (except
color-suppressed $B$ decays such as $B^0 \to \bar{D}^0\pi^0$),
which we call $\BB$ background in this paper,
have a non-peaking $\Mbc$ distribution
which is modeled by another ARGUS function.
Rare $B$ decays, i.e.\ $B$ decays through $b \to s$ and $b \to u$
transitions (charmless $B$ decay) and color-suppressed $B$ decays,
are not negligible and have peaks at the $B$ mass
in the $\Mbc$ distribution. The sum of these distributions
is modeled by a Gaussian plus an ARGUS function.
The shape of the distributions is determined by
a corresponding Monte Carlo (MC) sample.

The $\Mbc$ distribution of the signal component is also modeled
by the sum of a Gaussian and an ARGUS function.
All parameters are determined from the $B \to X_s\gamma$ signal MC
simulation,
except that the mean of the Gaussian is extracted
from a fit to $B \to D\pi$ decays, described below.
Our nominal signal MC contains a $B \to K^*(892)\gamma$
component and an inclusive $\bsgamma$ component
with $\MXs > 1.15~\GeV/c^2$.
The ratio of the two components is
based on
the branching fraction for $B \to K^*(892)\gamma$ measured by
Belle~\cite{Belle:kstgamma}
and the PDG value of $\BF(B \to X_s\gamma)$~\cite{Hagiwara:2002pw}.
The $X_s$ system of $\bsgamma$ is modeled
as an equal mixture of $s\bar{d}$ and $s\bar{u}$
quark pairs, and is hadronized using JETSET~\cite{Sjostrand:1994yb}.
The $\MXs$ spectrum is fitted to the model by Kagan and
Neubert~\cite{Kagan:1998ym}
with the $b$ quark mass parameter $m_b = 4.75~\GeV/c^2$.

Before discussing the extraction of $\ACP$,
we elaborate
on the signal modeling.
We create alternative signal MC samples
in which the $K^*(892)\gamma$ fraction
is varied by $\pm 1\sigma$ or the $b$ quark mass parameter is
varied by $\pm 0.15~\GeV/c^2$.
We account for the uncertainty in the hadronization
process by preparing an alternate MC sample wherein
$B$ candidates with $\MXs > 1.15~\GeV/c^2$ are
selected or discarded at random to match the fractions
of such candidates observed
in the data having between one and four pions.
We prepare other such samples wherein the proportion of
selected $B$ candidates without a neutral kaon or pion matched the value
seen in data; the correction is modest for all modes
except $K\pi\gamma$ (which contributes $32\%$ in the MC
but $12\%$ after correction).
The systematic error estimate based on these samples is described below.

In the pseudo-reconstruction analysis of $B \to X_s\gamma$,
the flavor of the $B$ meson is
``self-tagged''
when the net charge of the $X_s$ system is non-zero
or the $X_s$ system contains an odd number of charged kaons.
Otherwise the flavor is ``ambiguous.''
Because we have three possibilities for the flavor tag,
there are three ways to tag the flavor incorrectly:
$w_1$ is the probability of classifying a self-tagged event
as a self-tagged event of the opposite flavor;
$w_2$ is the probability to classify an ambiguous event
as self-tagged;
$w_3$ is the probability to classify a taggable event as ambiguous.
For example, $w_2$ refers to the case
when a $B^0 \to \KS\pi^+\pi^-\gamma$ event
is tagged as a $\KS\pi^+\pi^0\gamma$ event,
and $w_3$ refers to the case
when a $B^+ \to \KS\pi^+\pi^0\gamma$ event is tagged as
a $\KS\pi^+\pi^-\gamma$.
The effect of the flavor dependence of the wrong tag fractions
is negligibly small, and is included in the systematic error from the
$B \to D\pi$ study described later.

The formula to calculate $\ACP$ is then
$\ACP = D \ACPraw$ with
the dilution factor
$D = (1-w_2-w_3)/[(1-w_2)(1-2w_1-w_3)]$
and the raw asymmetry
$\ACPraw = (N_- - N_+)/[N_- + N_+ - (w_2/(1-w_2)) N_0]$,
where $N_-$ ($N_+$) is the number of events tagged as
originating from a $b$ ($\bar{b}$) quark,
and $N_0$ is the number of events classified as ambiguous.

The three wrong tag fractions are estimated using signal MC,
and thus are model dependent.
We calculate the wrong tag fractions for the collection of signal MC
samples described earlier,
and extract the systematic errors in the wrong tag fractions
from the changes in $w_i$ as the model is changed.
We find $w_1 = \wrongtagOne$, $w_2 = \wrongtagTwo$
and $w_3 = \wrongtagThree$, resulting in a dilution
factor of $D = \dilution$.

\begin{figure*}
 \includegraphics[scale=.54]{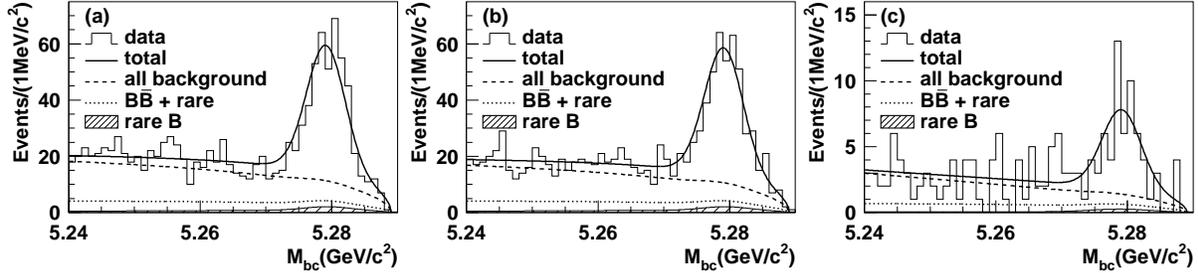} 
 \caption{\label{fig:mbcfit}%
 $\Mbc$ distributions for (a) events
 tagged as $b$, (b) events tagged as $\bar{b}$ and
 (c) events classified as ambiguous.
 Fit results are overlaid.}
\end{figure*}

Figure~\ref{fig:mbcfit} shows
the $\Mbc$ distributions for events that
are classified as $b$-tagged,
$\bar{b}$-tagged and ambiguous, respectively.
The distributions are fitted with signal, $\qq$,
$\BB$ and rare $B$ components.
We assume the shapes of the three components are
common for $b$- and $\bar{b}$-tagged classes,
distinct from those for the ambiguous class.
In the fit, the numbers of events from the $\BB$ and rare $B$ background
in the signal region are fixed to be
$\YBBFixedTagged$ ($\YBBFixedAmbiguous$) and
$\YrareBFixedTagged$ ($\YrareBFixedAmbiguous$)
for the common $b$- and $\bar{b}$-tagged classes (ambiguous class)
using the MC prediction,
while the normalization of the $\qq$ component is allow to float.
Signal yields are obtained by integrating the signal
Gaussian and ARGUS functions in the signal region.
We find the signal yield in each class
to be $N_- = \YieldMinusNosyst$, $N_+ = \YieldPlusNosyst$
and $N_0 = \YieldZeroNosyst$,
resulting in $\ACP = \ACPnosyst$.

The systematic errors are studied using a $B \to D\pi$
control sample. $D$ mesons are reconstructed in the same final states
as $X_s$ except for the final states to which $D$ cannot decay.
Only a mild requirement on the invariant mass of the reconstructed
$D$ mesons of less than $1.9~\GeV/c^2$ is applied,
in order to allow cross-feeds between different $D$ decay channels.
In the calculation of $\Mbc$,
the primary pion momentum is rescaled
in the same way as the photon in the $B \to X_s\gamma$ reconstruction.

In order to check the validity of the signal $\Mbc$ shape obtained
from MC, we compare the $\Mbc$ distribution for $B \to D\pi$ data and MC.
The data $\Mbc$ distribution is fitted with the sum of a
Gaussian and an ARGUS function plus
$\qq$ and $\BB$ distributions.
The $\qq$ and $\BB$ distributions
are obtained from off-resonance data and MC simulation, respectively,
where each normalization is scaled according to the luminosity.
From this procedure, we determine the mean of the Gaussian
used to model the signal $\Mbc$ distribution.
We find that the ratio of the Gaussian and ARGUS function
agrees between data and MC within the statistical error of
$\GaussLambdaAdderrratio$.

The systematic error on $\ACP$ due to the fitting procedure
is estimated by varying the value of each fixed parameter
by $\pm 1\sigma$ and
extracting new signal yields and $\ACP$ for each case.
We assign an additional $\GaussLambdaAdderrratio$ error
obtained from $B \to D\pi$ to the ratio of the Gaussian and ARGUS function.
The ratio is also varied separately for $N_+$ and $N_-$
to take into account the possible difference
of sub-decay modes between $b$ and $\bar{b}$.
We also use the signal shape parameters obtained from each of the signal
MC samples described earlier and extract $\ACP$ for each case.
We set the normalization of either the $\BB$ or rare $B$
backgrounds to zero and to twice its nominal value
to account for its uncertainty.
A $50\%$ error is assigned to the Gaussian width and to the ratio of the
Gaussian and ARGUS function for the rare $B$ decays, to compensate for our
limited knowledge of their branching fractions.
The changes of $\ACP$ for each procedure
are added in quadrature, and are regarded as the systematic error.
We obtain a systematic error of $\ACPerrorFit$ due to the fitting procedure.

The $B \to D\pi$ sample is also used
to estimate the possible detector bias in $\ACP$.
We estimate $\ACP$ for $B \to D\pi$ using the same method as
for $B \to X_s\gamma$.
The flavor is determined from
that of the $D$ candidate.
From the MC study, we find the $B \to D\pi$ wrong tag fractions to be
$w_1 = \wrongtagOneDpi$, $w_2 = \wrongtagTwoDpi$,
and $w_3 = \wrongtagThreeDpi$, resulting in
a dilution factor of $D = \dilutionDpi$.
The signal yields are calculated by fitting the data $\Mbc$ distribution
with signal, $\qq$, and $\BB$ components with fixed $\BB$ normalization,
and are found to be $N_- = \YieldMinusDpi$, $N_+ = \YieldPlusDpi$
and $N_0 = \YieldZeroDpi$,
resulting in a $\CP$ ``asymmetry'' of $\ACP = \ACPDpi$.
We therefore
assign $\ACPerrorDetector$ as the systematic error due to detector bias.

In addition, we estimate the uncertainty due to possible asymmetries
in charmless $B$ decays. We divide these decays into two groups:
those for which $\ACP$ has been measured
($B \to K^*\eta$, $K\pi^0$ and $K\rho$),
corresponding to about half of the selected charmless $B$ events;
and all other decay modes. We assign
an $\ACP$ of $\pm 30\%$ to the first group
(corresponding to around 1$\sigma$ in each case),
and $\pm 100\%$ to the second,
assuming the signs of $\ACP$ are correlated.
The resulting contribution to the measured $\ACP$,
$\ACPerrorRareB$, is taken as a systematic error.

The systematic error on $\ACP$ due to the uncertainty of the wrong tag
fractions is found to be small.
The contribution of the $\bdgamma$ process is negligible,
because we require the existence of kaons in the final state.
This is confirmed by examining MC samples of
$B \to \rho\gamma$, $\omega\gamma$ and
inclusive $\bdgamma$ process~\cite{footnote:bdgamma};
we expect that the contribution of $\bdgamma$
to $\ACP$ is less than $\ACPmaxerrorbdgamma$.
The total systematic error on $\ACP$
is then calculated to be $\ACPsyst$ by
adding in quadrature the errors mentioned earlier.

\begin{figure}
 \includegraphics[scale=.7]{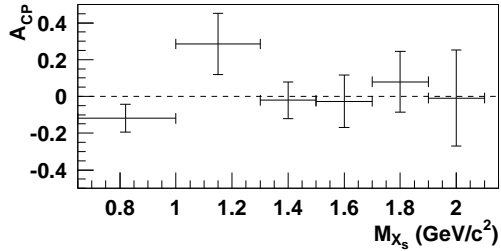}
 \caption{\label{fig:acpmxs}%
 $\ACP$ for each $\MXs$ bin.
 $\MXs$ is the reconstructed value of the mass of the $X_s$ system.
 The data points do not include any correction
 for the effect of smearing and mis-reconstruction.
 }
\end{figure}

In order to examine the $\MXs$ dependence of $\ACP$,
we divide the data sample into six bins of measured $\MXs$
and perform the $\Mbc$ fit for each bin.
Figure~\ref{fig:acpmxs} shows the $\ACP$ distribution
as a function of $\MXs$.
Here, the systematic error due to the detector bias
and possible $\CP$ asymmetry in charmless $B$ decays
is not included in the error.
We find that $\ACP$ is consistent with zero
for all $\MXs$ values
in the distribution.
From our MC study, about $2\%$ of the events reconstructed
with $\MXs < 2.1~\GeV/c^2$ are expected to have
a true $X_s$ mass greater than $2.1~\GeV/c^2$.

In conclusion,
the $\CP$ asymmetry of $B \to X_s\gamma$
for events with $\MXs < 2.1~\GeV/c^2$
is measured to be
\[ 
 \ACP(B \to X_s\gamma; \MXs<2.1~\GeV/c^2) = \ACPfinalSTR,
\]
consistent with no asymmetry.
This corresponds to $\ACPmin < \ACP < \ACPmax$ at the 90\% confidence level,
where we add the statistical and systematic errors in quadrature
and assume Gaussian errors.
The result can restrict the parameter space of new physics models
that allow sizable $\mathit{CP}$ asymmetry in $\bsgamma$~\cite{BabarACP}.

We thank the KEKB group for the excellent
operation of the accelerator, the KEK Cryogenics
group for the efficient operation of the solenoid,
and the KEK computer group and the NII for valuable computing and
Super-SINET network support.  We acknowledge support from
MEXT and JSPS (Japan); ARC and DEST (Australia); NSFC (contract
No.~10175071, China); DST (India); the BK21 program of MOEHRD and the
CHEP SRC program of KOSEF (Korea); KBN (contract No.~2P03B 01324,
Poland); MIST (Russia); MESS (Slovenia); NSC and MOE (Taiwan); and DOE
(USA).


\end{document}